%Paper: hep-ph/9410236
%From: PHRA0MG@TECHNION.TECHNION.AC.IL
%Date: Thu, 06 Oct 94 17:11:17 IST

%\magnification 1200
\vsize=21.0cm
\hsize=14.1cm
%\nopagenumbers
\baselineskip=6truemm
\def\ubar{\overline{u}}
\def\cbar{\overline{c}}
\def\tbar{\overline{t}}
\def\dbar{\overline{d}}
\def\sbar{\overline{s}}
\def\bbar{\overline{b}}
\def\fbar{\overline{f}}

\def\Abar{\overline{A}}
\def\Kbar{\overline{K}^0}
\def\Bbar{\overline{B}^0}
\def\Pbar{\overline{P}^0}
\def\tepsilon{\tilde{\epsilon}}
\def\bd{B^0}
\def\bs{B_s^0}
\def\bdb{\overline{B}^0}

\def\Dbar{\overline{D}^0}
\def\nubar{\overline{\nu}}
\def\Gammabar{\overline{\Gamma}}
\def\to{\rightarrow}

\def\PDG94{3}
\def\E731{4}
\def\NA31{5}

\def\PDG92{7}

\def\su3{39}

%\line{\hfil TECHNION-PH-94-54}
%\line{\hfil September, 1994}
%\null\vskip 2truecm

\centerline{\bf CP VIOLATION }
\vskip 2mm
\centerline{\it Michael Gronau}
\centerline{\it Department of Physics,
Technion -- Israel Institute of Technology}
\centerline{\it 32000 Haifa, Israel}

%\vskip 2.5cm
\vskip 1cm

%\magnification 1200
%\baselineskip=6truemm
%\noindent
%\centerline{\bf Abstract}
%\vskip 3mm
\noindent
We review the present status of the Standard Model of
CP violation, which is based on a complex phase in the
Cabibbo-Kobayashi-Maskawa
(CKM) matrix. So far CP violation has been observed only in $K^0-\Kbar$
mixing, consistent with a sizable phase. The implications of future CP
nonconserving measurements in $K$ and $B$ decays are discussed within the
model.
Direct CP violation in $K\to 2\pi$ may be observed in the near future, however
this would not be a powerful test of the model. $B$ decays provide a wide
variety of CP asymmetry measurements, which can serve as precise tests of the
Standard Model in cases where the asymmetry is cleanly related to phases of CKM
matrix elements.  Some of the most promising cases are discussed.
%\vskip 3cm
%\centerline{\it Invited talk presented at Neutrino 94,}
%\centerline{\it XVI International Conference on Neutrino Physics}
%\centerline {\it and Astrophysics, Eilat, Israel, May 29 - June 3, 1994}

%\vfill \eject
\vskip 1cm
\baselineskip=6truemm
\noindent
{\bf 1. INTRODUCTION}
\vskip 3mm

The history of the subject of CP violation is thirty years old. The first few
tens of $K_L\to\pi^+\pi^-$ events observed by Christenson, Cronin, Fitch and
Turlay [1] led the way to millions of events, in which long-lived
neutral kaons were observed to decay to both charged and neutral pions.
So far the neutral kaon system remains the only system in which CP
nonconservation has been measured. For three decades continuously
improving $K$ decay experiments verified the single fact that CP is violated in
$K^0-\Kbar$ mixing [2]. The parameter which describes this phenomenon is
$\epsilon$, the tiny CP impurity in the short- and long-lived kaon states. Its
present world-average value [3], $\vert\epsilon\vert=(2.26\pm 0.02)\times
10^{-3}$, is measured by $\vert\eta_{+-}\vert=\vert\eta_{00}\vert$, the $K_L$
to
$K_S$ ratio of decay amplitudes into charged and neutral two pion states.
The independent measurement of  $2{\rm Re}\epsilon$ by the charge asymmetry in
$K_L\to\pi\ell\nu$ [2] is consistent with this value and with the phase
measurement of $\eta_{+-}$.

The search for direct CP violation in the $K\to \pi\pi$ decay process
was the main purpose of two experiments operating for a lengthly period
starting in 1986 at Fermilab and at CERN. The results of these experiments,
looking for a difference between $\eta_{+-}$ and $\eta_{00}$, were
${\rm Re}(\epsilon'/\epsilon)=(7.4\pm 5.9)\times 10^{-4}$ (Fermilab E731
[4])  and $(23\pm 6.5)\times 10^{-4}$ (CERN NA31 [5]). This did not provide
unambiguous evidence for a nonzero effect.

CPT is a cherished symmetry of quantum field theory. It is quite
important to establish a high precision for this symmetry in the K system,
where it implies an equivalence between CP violation and the breaking
of time reversal symmetry. One such test [6], ${\rm Arg}(\eta_{+-})={\rm
Arg}(\eta_{00})=\tan^{-1}(-2\Delta m/\Delta\Gamma)$ (where $\Delta M$ and
$\Delta \Gamma$ are the neutral $K$ meson mass- and width-differences),
suffered
for a while from a two standard deviation discrepancy [7]. Recent experiments
at
Fermilab and at the Low Energy Antiproton Ring (LEAR) at CERN have, however,
measured somewhat smaller values for $\Delta m$ and for the phase of
$\eta_{+-}$, in very good agreement with this relation [8].

Future plans of experimental CP studies in strange particle decays [9] include
searching for direct CP violation in $K\to \pi\pi$, at CERN, at Fermilab and at
the Frascatti $\Phi$ factory. It is hoped to reach by 1996 a level of $10^{-4}$
in  ${\rm Re}(\epsilon'/\epsilon)$ and to improve the present precision of the
CPT test. Other decay processes, in which CP violation effects will be looked
for, include: $K\to 3\pi$, $K_L\to \pi^0 e^+e^-$, $K_L\to\pi^0\nu\nubar$ and
hyperon decays.

It seems, however, that the next decade of CP violation studies will be
dominated by the $B$ meson system. Very useful information, such as the large
$B^0-\Bbar$ mixing, measurements of relevant branching ratios,
and feasibility studies of rare decays which are important for CP violation,
was
already obtained from Doris II at DESY and from presently operating
facilities, mostly from CESR at Cornell, the LEP
accelerator at CERN and the Fermilab Tevatron. It is not entirely unlikely
that in the near future experiments at these accelerators will provide us with
first serious studies of CP violation in the $B$ system. In about four years
one
expects the starting operation of three large scale experiments dedicated to
thi
purpose, at two asymmetric $e^+e^-$ $B$ factories at SLAC and at KEK, and at
HERA using an internal target at the proton ring. By the end of this millennium
these facilities can provide measurements of CP asymmetries in a few $B$ decay
channels.  Further out in the future one may expect that the LHC at CERN will
provide us with the ultimate abundant production of $B$ mesons,
where a special $B$-physics program has the potential of high precision CP
studies.

The Standard Model [10] provides a suitable framework for understanding the CP
violation observed in the neutral $K$ meson mixing. The single source of CP
violation in the theory is a phase in the Cabibbo-Kobayashi-Maskawa
(CKM) matrix [11]. The only information about this phase comes from the
measured value of the CP impurity mixing parameter $\epsilon$. Thus, while this
single measurement can be accommodated in the CKM theory, it cannot test the
theory. The predictions of direct CP violation in strangeness-changing
processes, such as $K\to \pi\pi$ and other $K$ and hyperon decays involve
large theoretical uncertainties. These measurements are important for their own
sake, just to demonstrate CP violation outside $K^0-\Kbar$ mixing, however
due to theoretical uncertainties they cannot serve as powerful tests of the
Standard Model. On the other hand, the $B$ meson system provides a wide variety
of independent CP asymmetry measurements related to different sectors of the
CKM
matrix. Some of these asymmetries can be related to corresponding CKM phases in
manner which  is free of theoretical uncertainties. These
phases are fundamental parameters of the Standard Model, just as the electron
or the t-quark mass. Thus, it seems that CP violation in $B$
decays is due to become a fertile ground for testing the CKM symmetry breaking
mechanism.

In this theoretical review we discuss CP violation in the $K$ and $B$ meson
systems within the Standard Model. We begin in Section 2 by introducing the CKM
matrix. We summarize the available information on the magnitude of its
elements and on their CP violating phases. The general formalism of CP
nonconservation in neutral meson mixing is described in Section 3, where we
note
an important  difference between $K^0-\Kbar$ and $B^0-\Bbar$ mixing. Section 4
treats the $K$ meson system. We discuss our present theoretical understanding
of
$\epsilon$, the measured CP impurity in the neutral $K$ meson system, and the
theoretical uncertainties of calculating direct CP violation in $K\to\pi\pi$.
CP
violation in the $B$ meson system is studied in Section 5. We distinguish
between three kinds of CP nonconserving phenomena and discuss in turn
their predictions within the Standard Model: 1. CP violation in $B^0-\Bbar$
mixing; 2. CP violation which occurs when mixed neutral $B$ mesons decay to
states which are common decay products of $B^0$ and $\Bbar$; 3. Direct CP
violation in charged $B$ decays. Our focus is mainly on CP asymmetries which
can
be related to fundamental CKM phases in a manner which involves no, or very
small, theoretical uncertainties. We describes two different methods of
neutral $B$ flavor-tagging, which is needed for asymmetry measurements
in neutral $B$ decays.
Section 6 summarizes with a few concluding remarks.

This review is not supposed to be complete. Complementary discussions with
further references can be found in several previous reviews [12]. As mentioned,
we will only consider CP violation with $K$ and $B$ mesons. Other related
topics, such as the limit on the neutron electric dipole moment, the strong CP
problem and the baryon asymmetry in the universe , will not be dealt with
due to shortage of time and since they do not seem to have direct consequences
in the Standard Model. We will restrict our discussion to CP violating
predictions within the CKM Model. Studies of alternative possible mechanisms of
CP violation, with corresponding predictions, can be found elsewhere [12].

\vskip 0.5truecm
\noindent
{\bf 2. CP VIOLATION IN THE STANDARD MODEL: THE CKM MATRIX}
\vskip 3mm

In the standard model the $SU(3)_C\times
SU(2)_L\times U(1)_Y$  gauge group is spontaneously broken by the vacuum
expectation value of a single scalar Higgs doublet. CP violation occurs
in the interactions of the three families of left-handed quarks with the
charged
gauge boson:
$$
-\cal L =
\left(
\matrix{\ubar&\cbar&\tbar\cr}\right)
\left(\matrix{m_u&~&~\cr ~&m_c&~\cr ~&~&m_t\cr}\right)
\left(\matrix{u\cr c\cr t\cr}\right)
+ \left(
\matrix{\dbar&\sbar&\bbar\cr}\right)
\left(\matrix{m_d&~&~\cr ~&m_s&~\cr ~&~&m_b\cr}\right)
\left(\matrix{d\cr s\cr b\cr}\right)
$$
$$
+{g\over\sqrt{2}}
\left(
\matrix{\ubar&\cbar&\tbar\cr}\right)_L \gamma^{\mu} V
\left(\matrix{d\cr s\cr b\cr}\right)_L W_{\mu}^+ +...
\eqno(1)
$$
CP violation requires a complex Cabibbo-Kobayashi-Maskawa
[11] (CKM) mixing matrix
$V$. The quark mass terms exhibit a symmetry under phase redefinitions
of the six quark fields. This freedom leaves a single phase in $V$. The unitary
matrix $V$, which can be defined in terms of this phase ($\gamma$) and three
Euler-like mixing angles, is approximated for most practical purposes by the
following form:
$$
V\approx\left(
\matrix{1&\vert V_{us}\vert&\vert V_{ub}\vert e^{-i\gamma}\cr
-\vert V_{us}\vert&1&\vert V_{cb}\vert\cr
\vert V_{us}V_{cb}\vert-\vert V_{ub}\vert e^{i\gamma}&-\vert
V_{cb}\vert&1\cr}\right)~.\eqno(2)
$$

The measured values of the three mixing angles ($\sin\theta_{12}\equiv\vert
V_{us}\vert,~\sin\theta_{23}\equiv\vert V_{cb}\vert,~
\sin\theta_{13}$

\noindent
$\equiv\vert V_{ub}\vert$) have a hierarchical pattern in
generation space [13],
$$
\vert V_{us}\vert=0.220\pm 0.002~(\lambda)~,
$$
$$
\vert V_{cb}\vert=0.038\pm
0.005~({\cal O}(\lambda^2))~,
$$
$$
\vert V_{ub}\vert=0.0035\pm 0.0015~({\cal
O}(\lambda^3))~,\eqno(3)
$$
often characterized by powers of a parameter $\lambda\equiv\sin\theta_c=0.22$
[14].   This structure was used with unitarity to obtain the approximate
expressions of  the three $t$ quark couplings in $V$. It is amusing to note
that
the yet unmeasured value of $\vert V_{tb}\vert$ obtained from unitarity is the
most accurately known parameter of the mixing matrix.

Unitarity of $V$ can be
represented geometrically in terms of triangles, such as the one depicted in
Fig. 1 representing the relation
$$
V_{ud}V^*_{ub}+V_{cd}V^*_{cb}+V_{td}V^*_{tb}=0~.\eqno(4)
$$

\vskip 5.5truecm
\centerline{\rm Figure 1: The CKM unitarity triangle}
\vskip 0.5truecm

\noindent
The three angles of the unitarity triangle, $\alpha,~\beta$ and $\gamma$
(which appears as a phase in (2)), are rather badly known at present. Current
constarints, which depend on uncertainties in $K$- and $B$-meson hadronic
parameters, can be approximately summarized by the following ranges [15]:
$$
10^0\leq\alpha\leq 150^0,~~~5^0\leq\beta\leq 45^0,~~~20^0\leq\gamma\leq 165^0~.
\eqno(5)
$$
As we will show, certain CP asymmetries in $B$ decays are directly related to
these angles in a manner which is free of hadronic uncertainties, and can
provide a more precise determination for some of these fundamental parameters.

One can draw similar unitarity triangles describing the orthogonality of
other pairs of columns or rows of the CKM matrix. Knowledge of the angles of
all
these triangles, which can be related to CP asymmetries, suffice to determine
the entire matrix [16]. All such triangles have equal areas, however they
involve one side which is much shorter than the other two sides, and
consequently one of their angles is very tiny. This is in contrast to the
angles
$\alpha, \beta$ and $\gamma$ which are naturally large, since  all the three
sides of the unitarity triangle of Fig. 1 are of comparable (${\cal
O}(\lambda^3)$) magnitude. Thus, for instance, the neutral $K$ meson triangle,
built of elements $V_{qd}V^*_{qs}~(q=u,c,t)$, has two long sides (length
$\lambda$) and one extremely short side (length ${\cal O}(\lambda^5)$). This
explains why CP asymmetries in K decays, which are related to the tiny angle of
this triangle (${\cal O}(\lambda^4)$), are of order $10^{-3}$.

\vskip 0.5truecm
\noindent
{\bf 3. CP VIOLATION IN NEUTRAL MESON MIXING}
\vskip 3mm

The flavor states $P^0$ and $\Pbar$ ($P$ can be either a $K$ or a $B$
pseudoscalar meson) mix through the weak interactions to form the "Light" and
"Heavy" mass-eigenstates $P_L$ and $P_H$:
$$
|P_L\rangle = p|P^0\rangle  + q|\Pbar\rangle~,
$$
$$
|P_H\rangle = p|P^0\rangle  - q|\Pbar\rangle~.
\eqno(6)
$$
These states have masses $m_{L,H}$ and widths $\Gamma_{L,H}$.
The Hamiltonian eigenvalue equation (using CPT)
$$
\left(
\matrix{M-{i\over 2}\Gamma&M_{12}-{i\over 2}\Gamma_{12}\cr
M^*_{12}-{i\over 2}\Gamma^*_{12}&M-{i\over 2}\Gamma\cr}\right)
\left(\matrix{p\cr \pm q\cr}\right)=(m_{L,H}-{i\over 2}\Gamma_{L,H})
\left(\matrix{p\cr \pm q\cr}\right)\eqno(7)
$$
has the following solution for the mixing parameter
$q/p\equiv(1-\tilde{\epsilon})/ (1+\tilde{\epsilon})$:
$$
{q\over p}= \sqrt{{M^*_{12}-{i\over 2}\Gamma^*_{12}\over
M_{12}-{i\over 2}\Gamma_{12}}}
=-{2(M^*_{12}-{i\over 2}\Gamma^*_{12})\over \Delta m-{i\over 2}\Delta \Gamma}
{}~,\eqno(8)
$$
where $\Delta m\equiv m_H-m_L, \Delta\Gamma\equiv \Gamma_H-\Gamma_L$.
$M_{12}$ and $\Gamma_{12}$ describe respectively transitions from $P^0$ to
$\Pbar$ via virtual states and contributions from decay channels which are
common to $P^0$ and $\Pbar$.

The CP impurity parameter $\tilde{\epsilon}$
gives the mass-eigenstates in terms of states with well-defined CP
$$
|P_L\rangle={1\over\sqrt{1+\vert\tepsilon\vert^2}}(|P^0_1\rangle + \tepsilon
|P^0_2\rangle)~,
$$
$$
|P_H\rangle={1\over\sqrt{1+\vert\tepsilon\vert^2}}(|P^0_2\rangle + \tepsilon
|P^0_1\rangle)~,
$$
$$
|P_1\rangle={1\over\sqrt{2}}(|P^0\rangle + |\Pbar\rangle)~,
$$
$$
|P_2\rangle={1\over\sqrt{2}}(|P^0\rangle - |\Pbar\rangle)~.\eqno(9)
$$

$q/p$ has a phase freedom under redefinition of the phases of the
flavor states $P^0,~\Pbar$: $|P^0\rangle\to e^{i\xi} |P^0\rangle,~
|\Pbar\rangle\to e^{-i\xi} |\Pbar\rangle~\Rightarrow ~(q/p)\to
e^{2i\xi}(q/p)$. Thus the phase of $q/p$ can be rotated away and $|q/p|=1$
means CP conservation in $P^0-\Pbar$ mixing. The deviation of $|q/p|$ from
one measures CP violation in the mixing:
$$
1-|{q\over p}|\approx 2{\rm Re}\tepsilon~,\eqno(10)
$$
where ${\rm Re}\tepsilon$ is phase-convention independent. For convenience,
we will use the quark phase convention in which the CKM matrix (2) is written.

It is clear from eq.(8) that CP violation in neutral meson mixing is expected
to
be  small under two different circumstances:
$$
{\rm Arg} M_{12}\approx {\rm Arg}(-\Gamma_{12})~~~~(K ~{\rm meson})~,
$$
$$
|\Gamma_{12}| \ll |M_{12}|~~~~(B ~{\rm meson})~.\eqno(11)
$$
The first case applies to the neutral $K$ meson system and the second one - to
$B$ mesons. The different circumstances allude to the reason for the small and
theoretically uncertain CP violation in $K$ decays in constrast to the large
and theoretically clean CP violation in $B$ decays.
In $K$ decays $\Gamma_{12}$ is dominated by the $2\pi$ channel, the
amplitude of which involves (in the CKM phase convention) a very small phase
which is even smaller than the small phase of $M_{12}$. The calculation of both
phases involve hadronic uncertainties. On the other hand, the second
condition, which applies to the neutral $B$ meson system, says nothing about
phases of decay amplitudes which can be and in fact are large. The phase of
$q/p$, which can be approximated by the phase of $M^*_{12}$, helps in relating
the expected large CP asymmetries to pure CKM parameters.

\vskip 0.5truecm
\noindent
{\bf 4. THE $K$ MESON SYSTEM}
\vskip 3mm
\noindent
{\bf 4.1 CP Violation in $K^0-\Kbar$ Mixing}
\vskip 3mm

In the CKM phase convention $M_{12}$ obtaines a small imaginary contribution
from $t$ and $c$ quarks in the box-diagrams of Fig. 2, and $\Gamma_{12}$ has a
much smaller imaginary part from $K\to 2\pi$ (see Sec. 4.2).

\vskip 5.5truecm
\centerline{\rm Figure 2: Box diagrams for ${\rm Im}M_{12}(K)$}
\vskip 0.5truecm
\noindent
Thus we have
$$
2|M_{12}|=\Delta m_K\equiv m_L-m_S~,
$$
$$
2|\Gamma_{12}|=-\Delta\Gamma_K\equiv \Gamma_S-\Gamma_L~,\eqno(12)
$$
where we used the conventional notations for the long- and short-lived kaons.
We find
$$
\tepsilon_K\approx{i{\rm Im}M_{12}\over \Delta m_K-{i\over 2}\Delta\Gamma_K}
={{\rm Im}M_{12}\over \sqrt{2}\Delta m_K} e^{i\phi_K}~,\eqno(13)
$$
where $\tan\phi_K\equiv -2\Delta m_K/\Delta\Gamma_K,~\phi_K=(43.6\pm 0.2)^0$
[3]
${\rm Re}\tepsilon$ (which is phase-convention independent) is measured in
$K\to 2\pi$ and by the charge asymmetry in $K_L\to\pi\ell\nu$.
We note that in the CKM phase convention also
the imaginary part of $\tepsilon$ is  given, to a good approximation, by ${\rm
Im}\epsilon$ as measured in $K\to 2\pi$ (see Sec. 4.2).

The calculation of ${\rm Im}M_{12}$ uses Fig. 2 with QCD corrections to
obtain the following expression for $\tepsilon$ [17]:
$$
|\tepsilon|\approx 60B_Kf(m_t,m_c,\eta_q,S_{ij})(S_{12}S_{23}S_{13})\sin
\gamma~.\eqno(14)
$$
The numerical coefficient includes factors such as $\pi^2, G^2_F, m^2_W, \Delta
m_K/m_K$ and $f^2_K$. ~$B_K$ gives the hadronic matrix element of the
box diagram in terms of the vacuum insertion value. A possible range
of values for this parameter is probably $B_K=0.8\pm 0.2$ [15]. The function
$f$
[17] involves the $c$ and $t$ quark masses, calculable QCD correction factors
$\eta_q$ and the quark mixing angles, $S_{12}\equiv |V_{us}|,~S_{23}\equiv
|V_{cb}|,~S_{13}\equiv |V_{ub}|$. The value of $f$ for allowed parameters
can range from 1 to 5. Finally, as any CP violating quantity, $|\tepsilon|$ is
proportinal to twice the area of the unitarity triangle,
$S_{12}S_{23}S_{13}\sin\gamma$, which can obtain values in the range $(1.5 -
4.5)\times 10^{-5}\sin\gamma$. We see that the experimental value
$|\tepsilon|=2.26\times 10^{-3}$ can be naturally obtained for a sizable phase
i
the range $\sin\gamma\sim 0.1-1$. The prediction for $|\tepsilon|$ includes,
aside from present uncertainties in CKM parameters, also theoretical
uncertainties in hadronic matrix elements and (perhaps to a less degree) -
uncertainties in quark mass values and in QCD effects.

\vskip 5mm
\noindent
{\bf 4.2 Direct CP Violation in $K\to 2\pi$}
\vskip 3mm

The weak amplitudes of neutral $K$ mesons to charged and to neutral
two pion states can be decomposed into amplitudes of final states
with isospin $I=0, 2$:
$$
\langle\pi^+\pi^-|H_W|K^0\rangle=\sqrt{{2\over 3}}A_0 e^{i\delta_0}+
\sqrt{{1\over 3}}A_2 e^{i\delta_2}~,
$$
$$
\langle\pi^0\pi^0|H_W|K^0\rangle=-\sqrt{{1\over 3}}A_0 e^{i\delta_0}+
\sqrt{{2\over 3}}A_2 e^{i\delta_2}~,
$$
$$
\langle\pi^+\pi^-|H_W|\Kbar\rangle=\sqrt{{2\over 3}}A^*_0 e^{i\delta_0}+
\sqrt{{1\over 3}}A^*_2 e^{i\delta_2}~,
$$
$$
\langle\pi^0\pi^0|H_W|\Kbar\rangle=-\sqrt{{1\over 3}}A^*_0 e^{i\delta_0}+
\sqrt{{2\over 3}}A^*_2 e^{i\delta_2}~.\eqno(15)
$$
$\delta_I$ is the elastic phase shift for $\pi\pi$ scattering at the kaon
mass in an isospin $I$ channel. $A_I$ involves a weak CKM phase $\phi_I$,
which changes sign under charge-conjugation, $A_I=|A_I|e^{i\phi_I}$. Defining
$$
\eta_{+-}\equiv {\langle\pi^+\pi^-|H_W|K_L\rangle\over
\langle\pi^+\pi^-|H_W|K_S\rangle}~,~~~
\eta_{00}\equiv {\langle\pi^0\pi^0|H_W|K_L\rangle\over
\langle\pi^0\pi^0|H_W|K_S\rangle}~,
$$
one finds
$$
\eta_{+-}=\epsilon+\epsilon'~, ~~~~~\eta_{00}=\epsilon-2\epsilon'~,\eqno(16)
$$
where
$$
\epsilon=\tepsilon+i\tan\phi_0~,
$$
$$
\epsilon'={w\over\sqrt{2}}(\tan\phi_2-\tan\phi_0)e^{i(\delta_2-
\delta_0+{\pi\over 2})}~.\eqno(17)
$$
It is not difficult to see that ${\rm Re}\epsilon'$ measures CP
violation in direct $K\to 2\pi$ decays. That is, it gives the rate
asymmetry between the instantaneous $K^0$ and $\Kbar$ decay widths into $2\pi$.

$\epsilon$ and $\epsilon'$ are phase-convention independent.
The equality ${\rm Re}\epsilon={\rm Re}\tepsilon$ does not depend on
phase-convention, and ${\rm Im}\epsilon\approx {\rm Im}\tepsilon$ holds in the
CKM phase convention, where
as we shall see  $\tan\phi_0\ll |\epsilon|$.
$\epsilon'$ is suppressed by the measured $\Delta I=3/2$ to $1/2$ suppression
factor $w\equiv {\rm Re}A_2/{\rm Re}A_0=0.045$ [3], aside from involving the
phases $\phi_0,~\phi_2$ which are by themselves much smaller than $|\epsilon|$.
This is basically the origin of the small value of $\epsilon'/\epsilon$ in the
Standard model. The phase of $\epsilon'$, $\delta_2-\delta_0+\pi/2=
(48\pm 4)^0$ [3]
is approximately equal to $\phi_K=(43.6\pm 0.2)^0$, the phase of $\epsilon$
(see
Sec. 4.1). Therefore $\epsilon'/\epsilon$ is approximately real. Since this
ratio is at most of order $10^{-3}$, one expects the
equality ${\rm Arg}\eta_{+-}={\rm Arg}\eta_{00}= \phi_K$ to hold within a high
precision if CPT invariance is valid.

A calculation of $\epsilon'/\epsilon$ requires knowing the phases
$\phi_0,~\phi_2$. These can be estimated in the Standard Model using the tree
and penguin diagrams as described in Fig. 3. Whereas the tree operator has real
contributions to both $A_0$ and $A_2$, the penguin operator comes with a
complex CKM phase and contributes only to $A_0$.

\vskip 4.5truecm
\centerline{\rm Figure 3: Tree and penguin diagrams in $K\to 2\pi$}
\vskip 0.5truecm

\noindent
Thus, one finds $\phi_2=0$ and $\phi_0$ can be estimated from Fig. 3 to be
given   by
$$
\tan\phi_0\sim {{\rm Im}(V_{td}V^*_{ts})\over V_{ud}V^*_{us}}({P\over T})\sim
{\rm a~few}\times 10^{-4}({P\over T})~.\eqno(18)
$$
With $P/T\sim{\cal O}(1)$ this implies $\epsilon'/\epsilon\sim{\cal
O}(10^{-3})$.  An actual calculation of $\epsilon'/\epsilon$ is quite
complicated [18], and involves large theoretical uncertainties in hadronic
matrix elements of tree and penguin operators, on top of the experimental
uncertainties in CKM elements. With the heavy $t$ quark, additional electroweak
penguin amplitudes (in which the gluon in Fig. 3 is replaced by a photon and by
a $Z$ boson) lead to complex contributions to $A_2$, through which $\phi_2$
tends to cancel the $\phi_0$ term in $\epsilon'$. This enhances the uncertainty
in $\epsilon'/\epsilon$. Any value in the range from a few times $10^{-5}$ to
$10^{-3}$ seems to be possible. Measurement of a nonzero value for
$\epsilon'/\epsilon$ at a level of $10^{-4}$ would be an important observation
by itself, however it cannot provide a precise test of the Standard Model.

\vskip 0.5truecm
\noindent
{\bf 5. THE $B$ MESON SYSTEM}
\vskip 3mm
\noindent
{\bf 5.1 CP violation in $B^0-\Bbar$ Mixing}
\vskip 3mm

In the $B$ system one has $|\Gamma_{12}|\ll |M_{12}|$. $\Gamma_{12}$ is
given by the absorptive part of the box diagram, Fig. 4(a), arising from decay
channels which are common to $B^0$ and $\Bbar$. On the other hand, $M_{12}$ is
the dispersive part of the diagram, Fig. 4(b), governed by the $t$ quark mass.
Crudely speaking $|\Gamma_{12}/M_{12}|\sim m^2_b/m^2_t$. Thus CP violation
in  $B^0-\Bbar$ mixing is expected to be very small in the Standard
Model [19], $2{\rm Re}\epsilon_B\approx 1-|q/p|\sim {\cal O}(10^{-3})$. This
estimate, which is about the level of violation measured in the neutral $K$
meso
system, involves hadronic uncertainties and cannot provide a useful
quantitative test of the Standard Model. A much larger value of
${\rm Re}\epsilon_B$ would be evidence against the CKM mechanism.

\vskip 5.5truecm
\centerline{\rm Figure 4: Box diagrams of $\Gamma_{12}$ (a) and $M_{12}$ (b)}
\vskip 0.5truecm

CP violation in $B^0-\Bbar$ mixing is expected to show up as a charge asymmetry
in semileptonic decays to "wrong charge" leptons, namely leptons to which only
a
mixed neutral $B$ can decay:
$$
A_{SL}={\Gamma(\Bbar(t)\to\ell^+\nu X)-\Gamma(B^0(t)\to\ell^-\overline{\nu}X)
\over \Gamma(\Bbar (t)\to\ell^+\nu X)+\Gamma(B^0(t)\to\ell^-\overline{\nu}
X)}~.
\eqno(19)
$$
$B^0(t)~(\Bbar(t))$ is a time-evolving state, which is created as a
$B^0~(\Bbar)$
state at $t=0$. The asymmetry can be easily shown to be time-independent:
$$
A_{SL}={1-|q/p|^4\over 1+|q/p|^4}\approx 4{\rm Re}\epsilon_B~.\eqno(20)
$$
There exists already an experimental upper limit from CLEO
[20], $|{\rm Re}\epsilon_B|<45\times 10^{-3}$ ($90\%$ c.l.), which is about two
orders of magnitude above the Standard Model prediction. It will be extremely
difficult to observe an asymmetry at the level predicted by the model. However,
further efforts to improve this limit, for the semileptonic decays which have a
large branching ratio, is definitely worthwhile

Let us note in passing that while CP violation in the $B^0-\Bbar$ {\it mixing}
i
expected to be as small as about the one observed in $K^0-\Kbar$
mixing, the  asymmetries expected in neutral $B$ {\it decays} are much larger
than those of $K$ decays. Thus, when discussing neutral $B$ decay asymmetries
in
the following section we will take $|q/p|=1$ which is a very good
approximation.
In this approximation the mixing parameter is a pure phase
$$
{q\over p}\approx \sqrt{M^*_{12}\over M_{12}}\equiv e^{-2i\phi_M}=\cases
{e^{-2i\beta}~&for $\bd$~,\cr 1~&for $\bs$~.\cr}
\eqno(21)
$$
We will also assume $\Gamma_L=\Gamma_H$, which is a good
approximation, in particular for $\bd$ where it is expected to hold with an
accuracy better than $1\%$

\vskip 5mm
\noindent
{\bf 5.2 CP violation in decays of mixed $B^0-\Bbar$}
\vskip 3mm

\noindent
{\it 5.2.1 Time-dependent asymmetries in the general case}
\vskip 3mm

Consider the time-evolution of a state which is identified at time $t=0$ as a
$B^0$:
$$
t=0:~~~~|B^0\rangle ={e^{-i\phi_M}\over\sqrt{2}}(|B_L\rangle+|B_H\rangle)~.
\eqno(22)
$$
The time-evolutions of the states $B_{L,H}$ are given simply by their masses
and
by their equal decay width $\Gamma$: $|B_{L,H}(t=0)\rangle \to
|B_{L,H}(t)\rangle= \exp[-i(m_{L,H}-{i\over 2}\Gamma)t]|B_{L,H}(t=0)\rangle$.
Thus, the $B^0$ oscillates into a mixture of $B^0$ and $\Bbar$:
$$
t:~~~~~~|B^0(t)\rangle=e^{-i\overline{m}t}e^{-{\Gamma\over 2}t}[\cos({\Delta m
t\over 2})|B^0\rangle+ie^{-2i\phi_M}\sin({\Delta mt\over 2})|\Bbar\rangle]~,
\eqno(23)
$$
where $\overline{m}\equiv(m_H+m_L)/2,~\Delta m\equiv m_H-m_L$.
Now, assume that both $B^0$ and $\Bbar$ can decay to a common state $f$, with
amplitudes $A$ and $\Abar$, respectively. The time-dependent decay rate to $f$
of an initial $B^0$ and the corresponding rate for an initial $\Bbar$ are
$$
\Gamma(B^0(t)\to f)=e^{-\Gamma t}|A|^2[\cos^2({\Delta mt\over 2})+|\Abar/A|^2
\sin^2({\Delta mt\over 2})-{\rm Im}(e^{-2i\phi_M}\Abar/A)\sin(\Delta mt)]~,
$$
$$
\Gamma(\Bbar (t)\to f)=e^{-\Gamma t}|A|^2[|\Abar/A|^2\cos^2({\Delta mt\over 2})
+\sin^2({\Delta mt\over 2})+{\rm Im}(e^{-2i\phi_M}\Abar/A)\sin(\Delta mt)]~.
\eqno(24)
$$
In the special case that $f$ is an eigenstate of CP,
$CP|f\rangle=\pm|f\rangle$, CP violation is manifest when
$\Gamma(t)\equiv\Gamma(B^0(t)\to f)\ne \Gamma(\Bbar (t)\to
f)\equiv\overline{\Gamma}(t)$. The CP asymmetry is given
by [21]:
$$
Asym.(t)\equiv {\Gamma(t)-\Gammabar(t)\over \Gamma(t)+\Gammabar(t)}=
{(1-|\Abar/A|^2)\cos(\Delta mt)-2{\rm
Im}(e^{-2i\phi_M}\Abar/A)\sin(\Delta mt)\over 1+|\Abar/A|^2}~.
\eqno(25)
$$
The two terms in the numerator represent different sources of CP violation.
The first term
follows from CP violation in the direct decay of a neutral $B$
meson, whereas the second term is induced by $B^0-\Bbar$ mixing.

\vskip 5mm
\noindent
{\it 5.2.2 Decay to CP eigenstates dominated by a single CKM phase}
\vskip 3mm

Let us first consider the case of no direct CP violation, $|\Abar|=|A|$, in
whic
a single weak amplitude (or rather a single weak phase)
dominates the decay [22]. This is the case of a
maximal interference term in Eq.(24). Denoting the weak and strong phases by
$\phi_f$ and $\delta$, respectively, we have $A=|A|\exp(i\phi_f)\exp(i\delta),~
\Abar=\pm|\Abar|\exp(-i\phi_f)\exp(i\delta)$, and the asymmetry is given
simply by
$$
Asym.(t)=\pm\sin2(\phi_M+\phi_f)\sin(\Delta mt)~.
\eqno(26)
$$
The sign is given by $CP(f)$. The time-integrated asymmetry is
$$
Asym.= \pm\big({\Delta m/\Gamma\over 1+(\Delta m/\Gamma)^2}\big )\sin2(\phi_M+
\phi_f)~.
\eqno(27)
$$
That is, in this case {\it the CP asymmetry measures a CKM phase with no
hadronic uncertainty}. The integrated asymmetry in $\bd$ decays may be as large
as $(\Delta m/\Gamma)/[1+(\Delta m/\Gamma)^2]=0.47$.

The best example is the well-known and much studied case [23] of $\bd\to \psi
K_S$, for which a branching ratio of about $5\times 10^{-4}$ has already been
measured [24]. In this case
$\phi_M=\beta,~\phi_f={\rm Arg}(V^*_{cb}V_{cs})=0,~CP(\psi K_S)=-1$. Another
case is $\bd\to\pi^+\pi^-$, for which a combined branching ratio
$BR(\bd\to\pi^+\pi^-~{\rm and}~K^+\pi^-)=(2.3\pm 0.8)\times 10^{-5}$ has been
measured [25], with a likely solution in which the two modes have about equal
branching ratios. In this case $\phi_f={\rm
Arg}(V^*_{ub}V_{ud})= \gamma,~CP(\pi^+\pi^-)=1$. Consequently one has in these
two cases    $$ Asym.(\bd\to\psi K_S;t)=-\sin2\beta\sin(\Delta mt)~,
$$
$$
Asym.(\bd\to\pi^+\pi^-;t)=-\sin2\alpha\sin(\Delta
mt)~.\eqno(28)
$$
In the case of decay to two pions the asymmetry obtains, however, corrections
from a second (penguin) CKM phase. This problem will be discussed below.

\vskip 5mm
\noindent
{\it 5.2.3 Decay to non-CP eigenstates}
\vskip 3mm
\noindent

Angles of the unitarity triangle can also be determined from neutral B decays
to states $f$ which are not eigenstates of CP [26]. This is feasible when
both a $B^0$ and a $\Bbar$ can decay to a final state which appears in only one
partial wave, provided that a single CKM phase dominates each of the
corresponding decay amplitudes.

The time-dependent rates for states which are $B^0$ or $\Bbar$ at $t=0$ and
decay at time $t$ to a state $f$ or its charge-conjugate $\fbar$ are given
by [27]:
$$
\Gamma_f(t)=e^{-\Gamma t}[|A|^2\cos^2({\Delta mt\over 2})+|\Abar|^2\sin^2
({\Delta mt\over 2})+|A\Abar|\sin(\Delta\delta+\Delta\phi_f+2\phi_M)\sin(\Delta
mt)]~,
$$
$$\Gammabar_f(t)=e^{-\Gamma t}[|\Abar|^2\cos^2({\Delta mt\over
2})+|A|^2\sin^2 ({\Delta mt\over
2})-|A\Abar|\sin(\Delta\delta+\Delta\phi_f+2\phi_M)\sin(\Delta mt)]~,
$$
$$
\Gamma_{\fbar}(t)=e^{-\Gamma t}[|\Abar|^2\cos^2({\Delta mt\over
2})+|A|^2\sin^2 ({\Delta mt\over
2})-|A\Abar|\sin(\Delta\delta-\Delta\phi_f-2\phi_M)\sin(\Delta mt)]~,
$$
$$
\Gammabar_{\fbar}(t)=e^{-\Gamma t}[|A|^2\cos^2({\Delta mt\over
2})+|\Abar|^2\sin
({\Delta mt\over 2})+|A\Abar|\sin(\Delta\delta-\Delta\phi_f-2\phi_M)\sin(\Delta
mt)]~.\eqno(29)
$$
Here $\Delta\delta,~(\Delta\phi_f)$ is the difference between the strong
(weak) phases of $A$ and $\Abar$. The four rates depend on four
unknown quantities, $|A|,~|\Abar|,~\sin(\Delta\delta+\Delta\phi_f+2\phi_M),~
\sin(\Delta\delta-\Delta\phi_f-2\phi_M)$. Measurement of the rates allows a
determination of the weak CKM phase $\Delta\phi_f+2\phi_M$ apart from a
two-fold ambiguity [26].

There are two interesting examples to which this method can be applied. In the
first case, $\bd\to\rho^+\pi^-$, one must neglect a second contribution of a
penguin amplitude, a problem which will be addressed in the following
subsection. Assuming for a moment that tree diagrams, shown in Figs. 5(a),
5(b),
dominate $A$ and $\Abar$, one can measure in this manner the angle $\alpha$,
since in this case $\Delta\phi_f+2\phi_M=2(\gamma+\beta)=2(\pi-\alpha)$. A
second case, which may be used to measure $\gamma$, is $\bs\to D^+_s K^-$,
in which only one amplitude contributes to $A$ and another amplitude - to
$\Abar$.
\vfil\eject

{}~~
\vskip 5.0truecm
\centerline{\rm Figure 5: Diagrams of $\bd\to\rho^+\pi^-$ (a) and
$\bdb\to\rho^+\pi^-$ (b)}
\vskip 0.5truecm

\vskip 5mm
\noindent
{\it 5.2.4 Corrections from penguin amplitudes}
\vskip 3mm

A crucial question is, of course, how good is the assumption of a single
dominant CKM phase, which is needed for a clean determination of an angle
of the  unitarity triangle. One may try to answer this question experimentally
by looking for an extra $\cos(\Delta mt)$ term in the time-dependent asymmetry
o
Eq.(25) which describes CP violation in the direct decay of $B^0$. There is,
however, the danger that this term will be unobservably small, just because
the final state interaction phase difference happens to be small. The effect of
second amplitude on the coefficient of $\sin(\Delta mt)$, which is proportional
to the cosine of this phase-difference [21], may still be large. This will be
demonstrated below for $\bd\to\pi^+\pi^-$.

In a wide variety of decay processes there exists
a second amplitude due to ``penguin" diagrams [28] in addition to the usual
``tree" diagram. In general, the new contribution becomes more disturbing when
the process involves a stronger CKM-suppression.
The penguin-to-tree ratio of amplitudes is proportinal to the ratio of the
corresponding CKM factors and to a QCD fator $(\alpha_s(m^2_b)/12\pi){\rm ln}
(m^2_t/m^2_b)$. This ratio may be estimated for a given process. A few examples
of final states in $\bd$ decays,
with different levels of CKM suppression, are [21]:
$$
{{\rm Penguin}\over {\rm Tree}}= \cases
{10^{-3}~& $\psi K_S$~,\cr 0.05~& $D^+D^-~(D^{*+}D^-)$~,\cr 0.20~&
$\pi^+\pi^-~(\rho^+\pi^-)$~,\cr {\cal O}(1)~& $K_S\pi^0$~.} \eqno(30)
$$
These numbers represent quite crude estimates, since there exists no reliable
method to calculate hadronic matrix elements of penguin operators. One way to
obtain information about these matrix elements would be to measure pure penguin
processes, such as $\bd\to\phi K_S$. Another way will be mentioned when
discussing charged $B$ decays.

We see from Eqs.(30) that the decay $\bd\to \psi K_S$ remains a pure case,
with less than $1\%$ corrections, also
in the presence of penguin contributions. On the other hand, penguin effects on
the CP asymmetry of $\bd\to\pi^+\pi^-$ may be substantial. This is demonstrated
in Fig. 6, taken from Ref. 29, which shows the coefficient of the $\sin(\Delta
mt)$ term in the asymmetry as function of the angle $\alpha$ for a zero final
state interaction phase difference. The range of values comes from taking the
ratio (Penguin/Tree) to be anywhere between 0.04 and 0.20. For a ratio of
0.20, an asymmetry as large as 0.40 can possibly be measured even when
$\sin(2\alpha)=0$.

\vskip 5.5truecm
\centerline{\rm Figure 6: Asymmetry in $\bd\to\pi^+\pi^-$ as function of
$\alpha$}

\vskip 5mm
\noindent
{\it 5.2.5 Removing penguin corrections in $\bd\to\pi^+\pi^-$}
\vskip 3mm

It is possible to disentangle the penguin contribution in $\bd\to\pi^+\pi^-$
fro
the tree-dominating asymmetry by measuring also the rates of
$B^+\to\pi^+\pi^0$ and $\bd\to\pi^0\pi^0$. The method [30] is based on the
observation that the two weak operators contributing to the three
isospin-related processes have different isospin properties just as in $K\to
2\pi$. Whereas the tree operator is a mixture of $\Delta I=1/2$ and $\Delta
I=3/2$, the penguin operator is pure $\Delta I=1/2$. Denoting the  physical
amplitudes of $B\to \pi^+\pi^-, \pi^0\pi^0, \pi^+\pi^0$  by the charges of the
two corresponding pions, one finds from an isospin decomposition
$$
{1\over\sqrt{2}}A^{+-}=A_2-A_0~,~~~A^{00}=2A_2+A_0~,~~~A^{+0}=3A_2~,
\eqno(31)
$$
where $A_0$ and $A_2$ are the amplitudes for a $\bd$ or a $B^+$ to decay into a
$\pi\pi$ state with $I=0$ and $I=2$, respectively. This yields the complex
triangle relation
$$
{1\over\sqrt{2}} A^{+-} + A^{00} = A^{+0}~.\eqno(32)
$$
There is a similar triangle relation for the charge-conjugated processes:
$$
{1\over\sqrt{2}}{\overline{A}}^{+-}
+ {\overline{A}}^{00} = {\overline{A}}^{-0}~.\eqno(33)
$$
Here, ${\overline{A}}^{+-}$, ${\overline{A}}^{00}$, and
${\overline{A}}^{-0}$ are the amplitudes for the processes
$\bdb\to\pi^+\pi^-$, $\bdb\to\pi^0\pi^0$, and $B^-\to\pi^-\pi^0$,
respectively. The ${\overline{A}}$ amplitudes are obtained from the $A$
amplitudes by simply changing the sign of the CKM phases (the strong phases
remain the same).

The crucial point in the analysis is that the "tree" contribution to $A_2$
has a well-defined weak phase, which is given by the angle $\gamma$ of the
unitarity  triangle. (The electroweak penguin contribution to $A_2$ is
negligible).
$$
A_2=\vert A_2\vert e^{i\delta_2}e^{i\gamma}~,~~~
{\overline{A}}_2=\vert A_2\vert e^{i\delta_2}e^{-i\gamma}~.
\eqno(34)
$$
where $\delta_2$ is the $I=2$ final-state-interaction phase.
It is convenient to define $\tilde{A}=\exp(2i\gamma)\overline{A}$
so that $\tilde{A_2}=A_2$ and $\tilde{A}^{-0}=A^{+0}$. The two complex
triangles representing Eqs.
(32)(33) (where $\overline{A}$ is replaced by
$\tilde{A}$)  are shown in Fig. 7. They have a common base (CP is conserved in
$B^+\to\pi^+\pi^0$); however the length of their corresponding sides are
different. That is, CP is violated in $\bd\to\pi^+\pi^-$
and in $\bd\to\pi^0\pi^0$.

\vskip 5.5truecm
\centerline{\rm Figure 7: Isospin triangles of $B\to\pi\pi$}
\vskip 0.5truecm

The six sides of the two triangles are measured by the decay rates of $B^{\pm}$
and by the time-integrated rates of $\bd~(\bdb)$. This determines the two
triangles within a two-fold ambiguity; each triangle may be turned
up-side-down. The coefficients of the
$\sin(\Delta mt)$ term in $\bd\to\pi^+\pi^-$ measures the quantities
$$
{\rm Im}\thinspace(e^{-2i(\beta+\gamma)}{\tilde
{A}^{+-}\over A^{+-}})={\vert\tilde{A}^{+-}\vert\over\vert
A^{+-}\vert}\sin(2\alpha+\theta_{+-}),\eqno(35)
$$
where $\theta_{+-}$ is obtained from Fig. 7. (This angle vanishes in the
absence
of the penguin correction). This determines the angle $\alpha$.

The application of this method in asymmetric $e^+e^-$ $B$-factories
[31] is likely to suffer from a very small branching ratio of
$\bd\to\pi^0\pi^0$ (which is expected to be color-suppressed) and from the
difficulty of observing two neutral pions. Of course, if the penguin
term is small, its effect on the asymmetry of $\bd\to\pi^+\pi^-$ will be small.
When discussing charged $B$ decays we will mention how these decays can tell us
something about the magnitude of the penguin term in $\bd\to\pi\pi$.

A similar isospin analysis
was carried out for other decays in which penguin
amplitudes are involved [32]. In general, the precision of determining a
CKM phase becomes worse when a larger number of amplitudes must be related.
Also a few ambiguities show up in this case. In the case of
$\bd\to\rho\pi$ (and $B^+\to\rho\pi$) five physical decay amplitudes appear.
In this case the ambiguity can be resolved if a full Dalitz plot analysis
can be made for the three pion final states [33].

\vskip 0.5truecm
\noindent
{\bf 5.3. CP Violation in charged $B$ decays}
\vskip 3mm

\noindent
{\it 5.3.1 A theoretical difficulty}
\vskip 3mm

The simplest manifestations of $CP$ violation are different partial decay
widths for a particle and its antiparticle into corresponding decay modes.
Consider a general decay $B^+\to f$ and its charge-conjugate process
$B^-\to\fbar$. In order that these two proceses have different rates, two
amplitudes ($A_1, A_2$) must contribute, with different CKM phases ($\phi_1 \ne
\phi_2$) and different final state interaction phases ($\delta_1\ne\delta_2$):
$$  A(B^+\to f)~=~\vert A_1\vert  e^{i\phi_1}e^{i\delta_1}~+ ~\vert
A_2\vert e^{i\phi_2}e^{i\delta_2}~,
$$
$$
{}~~~~~\Abar (B^-\to \fbar)~=~\vert A_1\vert  e^{-i\phi_1}e^{i\delta_1}~+
{}~\vert
A_2\vert e^{-i\phi_2}e^{i\delta_2}~,
$$
$$
{}~~~~~~~~~\vert A \vert^2-\vert\Abar\vert^2=2\vert A_1
A_2\vert\sin(\phi_1-\phi_2)\sin(\delta_1-\delta_2)~. \eqno(36)
$$
The theoretical difficulty of relating an asymmetry in charged $B$ decays to a
pure CKM phase follows from having two unknowns in the problem: The
ratio of amplitudes, $\vert A_2/A_1\vert$, and the final state phase
difference,
$\delta_2-\delta_1$. Both quantities involve quite large theorertical
uncertainties.

This is demonstrated in Fig. 8, which describes the two amplitudes $A_1$ and
$A_2$ for $B^+\to K^+\pi^0$, given by the ``penguin"  and ``tree"
diagrams, respectively.

\vskip 5.5truecm
\centerline{\rm Figure 8: Penguin (a) and tree (b) diagrams in $B^+\to
K^+\pi^0$}
\vskip 0.5truecm

\noindent
In this case $\phi_1=\pi,~\phi_2=\gamma$. A few calculations of the
asymmetry in this process exist [34], based on model-dependent estimates of the
tree-to-penguin ratio of amplitudes and of the strong phase difference.
The strong phase includes a phase
due to the absorptive part of the physical $c\cbar$ quark pair in the penguin
diagram, which may be viewed as describing rescattering processes such as
$B\to\overline{D} D_s\to K\pi$. All such model-dependent calculations
involve large theoretical uncertainties.

\vskip 5mm
\noindent
{\it 5.3.2 Measuring $\gamma$ in $B^{\pm}\to D^0 K^{\pm}$}
\vskip 3mm

The decays
$B^{\pm}\to D^0_1(D^0_2) K^{\pm}$ and a few other processes of this
type provide a unique case [35], in which one can measure separately the
magnitudes of the two contributing amplitudes, and thereby determine the CKM
phase  $\gamma$. $D^0_1(D^0_2)=(D^0+(-)\Dbar)/\sqrt{2}$ is a CP-even (odd)
state, which is  identified by its CP-even (odd) decay products. For
instance, the states $K_S\pi^0,~K_S\rho^0,~K_S \omega,~K_S \phi$ identify a
$D^0_2$, while $\pi^+\pi^-,~ K^+K^-$ represent a $D^0_1$. The decay amplitudes
of the above two charge-conjugate processes can be written (say for $D^0_1$) in
the form
$$
\eqalign{ \sqrt{2}A(B^+\to D^0_1 K^+)~=~\vert
A_1\vert\exp(i\gamma)\exp(i\delta_1)~+~\vert
A_2\vert\exp(i\delta_2)~~,\cr \sqrt{2}A(B^-\to D^0_1 K^-)~=~\vert
A_1\vert\exp(-i\gamma)\exp(i\delta_1)~+~\vert
A_2\vert\exp(i\delta_2).\cr}\eqno(37)
$$
$A_1$ and $A_2$ are the two weak amplitudes, shown in Fig. 9(b) and 9(a),
respectively. Their CKM
factors $V^*_{ub}V^{~}_{cs}$ and $V^*_{cb}V^{~}_{us}$ are of comparable
magnitudes. Their weak phases are $\gamma$ and zero. Since $A_1$ leads to final
states with isospin $0$ and $1$, whereas $A_2$ can only lead to isospin $1$
states, one generally expects [36] $\delta_1\ne\delta_2$.

\vskip 5.5truecm
\centerline{\rm Figure 9: Diagrams decribing $B^+\to\Dbar K^+$ (a) and
$B^+\to D^0 K^+$ (b)}
\vskip 0.5truecm

As shown in Fig. 9, the two amplitudes on the right-hand-sides
of the first of Eqs. (37) are the amplitudes of $B^+\to D^0 K^+$ and
$B^+\to \Dbar K^+$, respectively. Similarly, the two terms in the second
equation describe the amplitudes of $B^-\to\Dbar  K^-$ and $B^-\to D^0 K^-$,
respectively. The flavor states $D^0$ and $\Dbar$ are
identified by the charge of the decay lepton or kaon. Thus one has:
$$
\eqalign{
\sqrt{2}A(B^+\to D^0_1 K^+)~=~A(B^+\to
D^0 K^+)~+~A(B^+\to \Dbar K^+),\cr \sqrt{2}A(B^-\to D^0_1 K^-)~=~A(B^-\to \Dbar
K^-)~+~A(B^-\to D^0 K^-).\cr}\eqno(38)
$$
Eqs. (38) can be described by two triangles in the complex plane  as shown in
Fig. 10.

\vskip 5.5truecm
\centerline{\rm Figure 10: Triangles describing Eqs.(38)}
\vskip 0.5truecm

The two triangles represent the complex $B^+$ and $B^-$ decay amplitudes. Note
that
$$
\eqalign{
A(B^+\to\Dbar K^+)=A(B^-\to D^0 K^-)~~~~~~~~~~,\cr
A(B^+\to D^0 K^+)=\exp(2i\gamma)A(B^-\to \Dbar K^-),\cr
\vert A(B^+\to D^0_1 K^+)\vert\ne\vert A(B^-\to D^0_1
K^-)\vert~~~~~~~~~.\cr}\eqno(39)
$$
This implies that CP is conserved in $B^{\pm}\to D^0(\Dbar) K^{\pm}$ but is
violated in  $B^{\pm}\to D^0_1 K^{\pm}$. In the last of Eqs.(39) we assumed
$\gamma\ne 0$, $\delta_1\ne\delta_2$. The asymmetry in the rates of
$B^{\pm}\to D^0_1 K^{\pm}$ depends on $\gamma$ and $\delta_2-\delta_1$;
clearly
$$
\eqalign{&
\vert A(B^+\to D^0_1 K^+)\vert^2-\vert A(B^- \to D^0_1 K^-)\vert^2 \cr &
=2\vert A(B^+\to\Dbar K^+)\vert\vert A(B^+\to D^0
K^+)\vert\sin(\delta_2-\delta_1)\sin\gamma. } \eqno(40)
$$

The procedure for obtaining $\gamma$ is straightforward. Measurements of the
rates of the above six proccesses, two pairs of which are equal, determine the
lengths of all six sides of the two triangles.
When the two triangles are formed, $2\gamma$ is the angle between
$A(B^+\to D^0 K^+)$ and $A(B^-\to \Dbar K^-)$. This determines the magnitude of
$\gamma$ within a two-fold ambiguity
related to a possible interchange of $\gamma$ and $\delta_1-\delta_2$. This
ambiguity may be resolved by carrying out this analysis for other decay
processes of the type $B^{\pm}\to D^0(\Dbar, D^0_{1(2)})X^{\pm}$, where
$X^{\pm}$ is any other state with the flavor quantum number of a $K^{\pm}$.

The feasibility of observing a CP asymmetry in $B^+\to D^0_{1(2)} K^+$ depends
on the branching ratios of the three related decay processes, and on the values
of the weak and strong phases. One may estimate $BR(B^+\to \Dbar K^+)\approx
2\times 10^{-4}$, using the corresponding measured Cabibbo-allowed branching
ratio  of $B^+\to\Dbar \pi^+$ [24]. The process
$B^+\to D^0 K^+$, in which the two quarks of the $c\sbar$ current
enter two different meson states, is likely to be "color-suppressed".
Color suppression has already been seen in $B\to D \pi$ [24]. If the same
suppression factor applies also to $B^+\to D^0 K^+$, then the branching ratio
of this process is at most at the level of $10^{-5}$. Using a value
of $5\times 10^{-6}$, the feasibility for observing a CP asymmetry in $B^+\to
D^0_{1(2)} K^+$ was studied [37] as function of $\gamma$ and
$\delta_2-\delta_1$, for a (symmetric) $e^+e^-\to\Upsilon(4S)$ $B$-factory with
an integrated luminosity of $20 fb^{-1}$. The discovery region was found
to cover a significant part of the ($\gamma, \delta_2-\delta_1$) plane. For
small final state phase differences the experiment is sensitive mainly to
values
of $\gamma$ around $90^0$. Large values of $\delta_2-\delta_1$ allow a
useful measurement of $\gamma$ in the range $50^0\leq\gamma\leq 130^0$.

Present experiments are reaching the level of being able to observe the first
Cabibbo suppressed decays $B\to DK$. The question of color-suppression in these
decays needs to be studied. It is possible that the final state phase
difference
$\delta_2-\delta_1$ is too small to allow a good measurement of $\gamma$ if
this angle is not around $90^0$. A recent study [38] generalized this method to
quasi-two-body decays  $B\to DK_i\to D K\pi$, where $K_i$ are excited kaon
resonance states with masses around 1400 MeV. The resonance effect
gives give
rise to large final state phases and thus enhances the CP asymmetry.

\vskip 5mm
\noindent
{\it 5.3.3 Using SU(3) to determine $\gamma$ from $B^+\to\pi K$ and
$B^+\to\pi\pi$}
\vskip 3mm

Flavor SU(3) symmetry can be used to relate $B$ decays to $\pi\pi,~\pi K$ and
$KK$ states [39]. Recently this idea was applied [40] jointly with the
dynamical
assumption that annihilation-like diagrams are small in these two-body
decays.  This assumption means that certain rescattering effects are small.
That
is, final states which are produced, for instance, by tree decay amplitudes
have
small rescattering amplitudes to states created by quark-antiquark
annihilation. This assumption is motivated by the high $B$ meson mass (compared
to $f_B$). It is supported by the experimental evidence for color-suppression
and for factorization in two body $B$ decays [24], two features which are
expected to be spoiled by large rescattering amplitudes. It was shown that the
assumption of negligible rescattering is equivalent to assuming that certain
final state phases are equal to others [41]. Neglecting such rescattering
effects leads to simple testable predictions, such as $A(B^0\to K^+ K^-)=0$,
and
to useful information about weak and strong phases. Here we wish to demonstrate
this idea through a rather simple case [42].

Consider the decay $B^+\to \pi^0 K^+$ for which the two contributing
amplitudes $A_1$ and $A_2$ are described in Fig. 8. The penguin amplitude
$A_1$ is related by isospin
to the amplitude of $B^+\to \pi^+K^0$, in which the annihilation contribution
is neglected, $A_1=A(\pi^+K^0)/\sqrt{2}$. The tree amplitude $A_2$
is related by SU(3) to the amplitude of $B^+\to\pi^+\pi^0$, which receives no
penguin contribution. Using factorization to introduce SU(3) breaking into this
relation, one has $A_2=(f_K/f_{\pi})|V_{us}/V_{ud}|A(\pi^+\pi^0)$. Thus one
obtains a simple relation between the three $B^+$ decay amplitudes:
$$
A(\pi^0 K^+)={1\over \sqrt{2}}A(\pi^+ K^0) + {f_K\over f_{\pi}}|{V_{us}\over
V_{ud}}|A(\pi^+\pi^0)~.\eqno(41)
$$
A similar relation holds among the corresponding $B^-$ decay amplitudes.
There are phase relations, $A(\pi^- \Kbar)=A(\pi^+ K^0),
A(\pi^-\pi^0)=\exp(-2i\gamma)A(\pi^+\pi^0)$, since the weak
phases of these amplitudes are $\pi$ and $-\gamma$, respectively. The two
relations among the $B^+$ and among the $B^-$ amplitudes are analogous to
Eqs.(38). They can be descrlibed by two triangles very similar to those of
Fig. 10. In the present case the two triangles share a common base given by
$A(\pi^- \Kbar)=A(\pi^+ K^0)$, and the angle between the sides describing
$A(\pi^-\pi^0)$ and $A(\pi^+\pi^0)$ is $2\gamma$. Measurements of the four
rates
into $\pi^0 K^+, \pi^0 K^-, \pi^+ K^0, \pi^+\pi^0$, suffices to determine
$\gamma$. CP violation is demonstrated by $A(\pi^0 K^+)\ne A(\pi^0 K^-)$. About
100 events of this mode, which is expected to have a branching ratio of about
$10^{-5}$, are needed to measure $\gamma$ to a statistical accuracy of $10^0$
[42].

This method is not as clean as the one using $B^{\pm}\to D^0K^{\pm}$ decays,
since it is based on certain dynamical assumptions. Also, it was recently noted
[43] that contributions from electroweak penguin diagrams can spoil the
relation
(41). Information about the effect of
these diagrams, of SU(3) breaking and of annihilation-like diagrams, and the
separate magnitudes of tree and penguin amplitudes, can be obtained from a
systematic study of all the possible $B$ decay modes to two light pseudoscalar
mesons [40]. Such a detailed study may then be used to evaluate the precision
to
which the weak phase can be determined.

\vskip 0.5truecm
\noindent
{\bf 5.4 Flavor-tagging of neutral $B$ mesons}
\vskip 3mm
\noindent
{\it 5.4.1 Tagging by the associated $B$ decay}
\vskip 3mm

In order to measure CP asymmetries in neutral $B$ decays one must identify the
flavor of the decaying meson at some reference time $t=0$. In a
$e^+e^-\to\Upsilon(4S)$ $B$-factory this is achieved [23] by observing a
lepton, or a cascade charged kaon from $B\to D\to K$, from the decay of the
othe
neutral $B$. Since at any time after production the two neutral $B$ mesons form
{\it coherent}  $C(B^0\Bbar)=-1$ EPR pair, the charge of the lepton serves to
"tag" the opposite flavor of the other $B$ at the time of  semileptonic decay.
Furthermore, the CP asymmetry is odd in the time-difference of the two decays,
and consequently asymmetric storage rings are required for an asymmetry
measurement.

A similar method of determining the flavor of neutral $B$ mesons
in high energy $e^+e^-$ or in hadronic collisions [44] uses as a "tag"
the lepton from a semileptonic decay of an associated $b$-meson or $b$-baryon.
The flavor is misidentified part of the time as a result of $B^0-\Bbar$ mixing.
The probability of misidentification and its effect on diluting the measured CP
asymmetry can be crudely estimated. Since the $B^0$ and $\Bbar$
are usually  produced with many other particles, it is commonly assumed that
the
are in an {\it incoherent} mixture.

\vskip 5mm
\noindent
{\it 5.4.2 Tagging by correlated charged pions}
\vskip 3mm

An alternative method of flavor identification [45] uses an expected
correlation
between the decaying neutral $B$ and a charged pion making a low-mass
$B-\pi$ system. There are two arguments for such a correlation.
The first argument is based on the existence of positive-parity ``$B^{**}$"
resonances, with $J^P=0^+,~1^+,~2^+$ and masses below about 5.8 GeV/$c^2$ [46].
Using Heavy Quark Symmetry, this mass value
is obtained from the corresponding observed
``$D^{**}"$ masses (2420, 2460 GeV/$c^2$). The $B^{**}$ resonances decay to
$B\pi$ and/or $B^*\pi$ mesons in $I=1/2$  states. That is, a $\pi^+$ will
accompany a $B^0$ and not a $\Bbar$. A similar method [47] has been used to
tag neutral charmed mesons, where the decays $D^{*+}\to D^0\pi^+,~D^{*-}\to
\overline{D}^0 \pi^-$ are kinematically allowed.
The second argument is that in $b$-quark fragmentation the leading pion carries
information about the flavor of the neutral $B$. A neutral $B$ meson
containing an initially produced $b$ quark is a $\Bbar$ which contains a
$\dbar$ quark. The next charged pion down the fragmentation chain must contain
a $d$, and hence must be a $\pi^-$. Similarly, a $B^0$ will be correlated
with a $\pi^+$.

The efficiency of this method depends on the degree of the correlation, which
can be studied in neutral $B$ decays to states of identified flavor, such as
$D^-\pi^+$ or $\psi K^{*0}$ (with $K^{*0}\to K^+\pi^-$). Usually $B$ mesons are
produced in an isospin-independent manner and one can find this correlation
using charged $B$ mesons as well. The time-dependent CP asymmetry measured with
this tagging method is diluted by the degree of correlation. Aside from using
the asymmetry to determine weak phases, it can also be used to test the
assumption that the produced $B^0$ and $\Bbar$ are incoherent with respect to
one another [48].

\vskip 0.5truecm
\noindent
{\bf 6. SUMMARY}
\vskip 3mm

The observed CP violation in  $K^0\Kbar$ mixing is successfully parametrized
in terms of a phase in the CKM matrix. This phase is largely unknown at
present.
Tests of the Standard Model of CP violation require more precise information
about magnitudes and phases of CKM elements. Future $K$
decay experiments may have the potential of measuring a nonzero value for
$\epsilon'/\epsilon$, thus confirming the expected phenomenon of direct CP
violation in $K$ decays. However, due to theoretical uncertainties this cannot
provide a precise test of the Standard Model and cannot cleanly determine CKM
parameters. On the other hand, measurements of certain CP asymmetries in $B$
decays can determine CKM phases in manners which are free of hadronic
uncertainties. At the very least, this will allow direct measurements of
these fundamental parameters. With an improved knowledge of the magnitudes of
CK
elements, this may eventually serve to overconstrain the CKM matrix. One would
hope to find some inconsistencies which could be clues for physics beyond the
Standard Model. Afterall, the observed baryon asymmetry in the universe seems
to
require other sources of CP violation [49].

\vskip 0.5truecm
\noindent
{\bf ACKNOWLEDGEMENTS}
\vskip 3mm

It is a pleasure to thank David Atwood, Gad Eilam, Oscar Hern\'andez, David
London, Alex Nippe, Jonathan Rosner, Amarjit Soni and Daniel Wyler for very
enjoyable collaborations on various topics presented here. This work was
supported in part by the United States - Israel Binational Science Foundation
under Research Grant Agreement 90-00483/3, by the German-Israeli Foundation of
Scientific Research and Development and by the Fund for Promotion of Research
at
the Technion.

\vskip 0.5truecm
%\vfil\eject
\noindent
{\bf REFERENCES}
\vskip0.3truecm

\item{1.} J. H. Christenson, J. W. Cronin, V. L. Fitch and R. Turlay, {\it
Phys. Rev. Lett.} {\bf 13} (1964) 138.
\item{2.} J. W. Cronin,  {\it Rev. Mod. Phys.} {\bf 53} (1981) 373; K.
Kleinknecht, in {\it CP Violation}, ed. C. Jarlskog (Singapore, World
Scientific, 1989), p. 41.
\item{3.} Particle Data Group, {\it Review of Particle Properties, Phys. Rev.}
{\bf D50} (1994) 1173.
\item{4.} E731 Collaboration, L. K. Gibbons {\it et al.}, {\it Phys. Rev.
Lett.}
{\bf 70} (1993) 1203.
\item{5.} NA31 Collaboration, G. D. Barr {\it et al.}, {\it Phys. Lett.} {\bf
B317} (1993) 233.
\item{6.} N. W. Tanner and R. H. Dalitz, {\it Ann. Phys. (N.Y.)} {\bf 171}
(1986) 463; T. Nakada, in {\it Proceedings of the XVI International Symposium
on Lepton and Photon Interactions}, Cornell University, August 10-15, 1993,
ed. P. Drell and D. Rubin (New York, AIP, 1994), p. 425.
\item{7.} Particle Data Group, {\it Review of Particle Properties, Phys. Rev.}
{\bf D45} (1992) 1.
\item{8.} E731 Collaboration, L. K. Gibbons {\it et al.}, {\it Phys. Rev.
Lett.} {\bf 70} (1993) 1199; E773 Collaboration, EFI 94-31, presented by B.
Schwingenheuer at the Fifth Conference on the Intersections of Particle and
Nuclear Physics, St. Petersburg, Florida, June 1994; CPLEAR Collaboration, C.
Yeche {\it et al.}, DAPNIA-SPP-94-18, June 1994.
\item{9.} B. Winstein and L. Wolfenstein, {\it Rev. Mod. Phys.} {\bf 65}
(1993) 1113.
\item{10.} S. L. Glashow, {\it Nucl. Phys.} {\bf 22} (1961) 579;
S. Weinberg, {\it Phys. Rev. Lett.} {\bf 19} (1967) 1264;
A. Salam, in {\it Elementary Particle Theory}, ed. N. Svartholm (Almqvist and
Wiksell, Stockholm, 1968).
\item{11.}  N. Cabibbo, {\it Phys. Rev. Lett.} {\bf 10}
(1963) 531;  M. Kobayashi and T. Maskawa, {\it Prog. Theor.~Phys.} {\bf 49}
(1973) 652.
\item{12.} {\it CP Violation}, {\it op. cit.}; Y. Nir and H. Quinn, {\it
Ann. Rev. Nucl. Part. Sci.} {\bf 42} (1992) 211; B. Winstein and L.
Wolfenstein,
Ref.9; J. L. Rosner, Enrico Fermi Institute Report EFI 94-25, presented at
PASCOS 94 Conference, Syracuse, NY, May 1994.
\item{13.} S. Stone, Syracuse University Report HEPSY-94-5, presented at
PASCHOS 94 Conference, {\it op. cit.}; A. Ali and D. London, CERN Report
CERN-TH.7398/94, August 1994.
\item{14.} L. Wolfenstein, {\it Phys. Rev. Lett.} {\bf 51} (1983) 1945.
\item{15.} A. Ali and D. London, Ref.13.
\item{16.} R. Aleksan, B. Kayser and D. London, {\it Phys. Rev. Lett.}
{\bf 73} (1994) 18.
\item{17.} T. Inami and C. S. Lim, {\it Prog. Theor.
Phys.} {\bf 65} (1981) 297; A. J. Buras, W. Slominski and H. Steger, {\it Nucl.
Phys.} {\bf B238} (1984) 529; {\bf B245} (1984) 369.
\item{18.} G. Buchalla, A. J. Buras and M. K. Harlander, {\it Nucl. Phys.} {\bf
B337} (1990) 313; J. Heinrich, E.A. Paschos, J. M. Schwarz and Y. L. Wu, {\it
Phys. Lett.} {\bf B279} (1992) 140; A. J. Buras, M. Jamin and M. E.
Lautenbacher, {\it Nucl. Phys.} {\bf 408} (1993) 209; R. Ciuchini, E. Franco,
G. Martinelli and L. Reina, {\it Phys. Lett.} {\bf B301} (1993) 263.
\item{19.} I. I. Bigi, V. A. Khoze, N. G. Uraltsev and A. I. Sanda, in {\it CP
Violation}, {\it op. cit.}, p. 175.
\item{20.} CLEO Collaboration, J. Bartelt {\it et al.}, {\it Phys. Rev. Lett.}
{\bf 71} (1993) 1680.
\item{21.} M. Gronau, {\it Phys. Rev. Lett.} {\bf 63} (1989) 1451.
\item{22.} A.B. Carter and A.I. Sanda, {\it Phys. Rev. Lett.} {\bf
45} (1980) 952; {\it Phys. Rev.} {\bf D23} (1981) 1567; I.I. Bigi and A.I.
Sanda
{\it Nucl. Phys.} {\bf B193} (1981) 85; {\bf B281} (1987) 41; I. Dunietz and
J.L
Rosner, {\it Phys. Rev.} {\bf D34} (1986) 1404.
\item{23.} {\it The Physics Program of a High-Luminosity
Asymmetric $B$-Factory at SLAC}, ed. D. Hitlin, SLAC-PUB-353 (1989) and
SLAC-PUB-373 (1991); {\it Physics Rationale for a $B$-Factory}, K. Lingel {\it
e
al.}, CLNS 91-1043 (1991); {\it Physics and Detector at KEK Asymmetric $B$
Factory}, K. Abe {\it et al.}, KEK Report 92-3 (1992).
\item{24.} CLEO Collaboration, M. S. Alam {\it et al.}, {\it Phys. Rev.} {\bf
D50} (1994) 43.
\item{25.} CLEO Collaboration, M. Battle {\it et al.}, {\it Phys. Rev. Lett.}
{\bf 71} (1993) 3922.
\item{26.} R. Aleksan, I. Dunietz, B. Kayser and F. Le Diberder,
{\it Nucl. Phys.} {\bf B361} (1991) 141; R. Aleksan, I. Dunietz and B. Kayser,
{\it Zeit. Phys.} {\bf C54} (1992) 653.
\item{27.} M. Gronau, {\it Phys. Lett.} {\bf B233} (1989) 479.
\item{28.} M. Gronau, Ref.21; D. London and R.D. Peccei, {\it Phys. Lett.}
{\bf B223} (1989) 257; B. Grinstein, {\it Phys. Lett.} {\bf B229} (1989) 280.
\item{29.} M. Gronau, {\it Phys. Lett.} {\bf B300} (1993) 163.
\item{30.} M. Gronau and D. London, {\it Phys. Rev. Lett.} {\bf 65} (1990)
3381.
\item{31.} R. Aleksan, A. Gaidot and G. Vasseur, in {\it $B$ Factories, The
Stat
of the Art in Accelerators, Detectors and Physics}, ed. D. Hitlin, SLAC-400
(1992), p. 588.
\item{32.} Y. Nir and H. Quinn, {\it Phys. Rev. Lett.} {\bf 67} (1991) 541; H.
J. Lipkin, Y. Nir, H. R. Quinn and A. E. Snyder, {\it Phys. Rev.} {\bf D44}
(1991) 1454; M. Gronau, {\it Phys. Lett.} {\bf B265} (1991) 389; L. Lavoura,
{\i
Mod. Phys. Lett.} {\bf A7} (1992) 1553.
\item{33.} H. R. Quinn and A. E. Snyder, {\it Phys. Rev.} {\bf D48} (1993)
2139.
\item{34.} L. L. Chau and H. Y. Cheng, {\it Phys. Rev. Lett.} {\bf 59} (1987)
958; N. Deshpande and J. Trampetic, {\it Phys. Rev.} {\bf D41} (1990) 2926; J.
M. Gerard and W. S. Hou, {\it Phys. Rev.} {\bf D43} (1991) 2909; H. Simma, G.
Eilam and D. Wyler, {\it Nucl. Phys.} {\bf B352} (1991) 367.
\item{35.} M. Gronau and D. Wyler, {\it Phys. Lett.} {\bf B265} (1991) 172. See
also M. Gronau and D. London, {\it Phys. Lett.} {\bf B253} (1991) 483; I.
Dunietz, {\t Phys. Lett} {\bf B270} (1991) 75.
\item{36.} I. I. Bigi and A. I. Sanda, {\it Phys. Lett.} {\bf B211} (1988) 213.
\item{37.} S. Stone, in {\it Beauty 93, Proceedings of
the First International Workshop on $B$ Physics at Hadron Machines}, Liblice
Castle, Melnik, Czech Republic, Jan. 18-22, 1993; ed. P. E. Schlein, {\it
Nucl. Instrum. Meth.} {\bf 333} (1993) 15.
\item{38.} D. Atwood, G. Eilam, M. Gronau and A. Soni, CERN Report
CERN-TH.7428/94, September 1994.
\item{39.} D. Zeppenfeld, {\it Zeit. Phys.} {\bf C8} (1981) 77; M. Savage and
M.
Wise, {\it Phys. Rev.} {\bf D39} (1989) 3346; {\it ibid.} {\bf 40} (1989)
3127(E); J. Silva and L. Wolfenstein, {\it Phys. Rev.} {\bf D49} (1994) R1151.
\item{40.}  M. Gronau, O. F. Hern\'andez, D. London, and J. L. Rosner, {\it
Phys.  Rev.} {\bf D50} (1994) Oct. 1; {\it Phys. Lett.} {\bf B333} (1994) 500;
see also A. J. Buras and R. Fleischer, Max-Planck-Institute Report
MPI-PhT/94-56
August 1994.
\item{41.} See also L. Wolfenstein, to be published in {\it Phys.
Rev.} {\bf D50} (1994).
\item{42.}  M. Gronau, J. L. Rosner and D. London, {\it
Phys. Rev. Lett.} {\bf 73} (1994) 21.
\item{43.} N. G. Deshpande and X-G He, University of Oregon Report OITS-553,
August 1994.
\item{44.} See e.g. {\it B Physics at Hadron Accelerators, Proceedings of a
workshop at Fermilab}, Nov., 1992, ed. J.A. Appel (Fermilab, Batavia, Il,
1992);
{\it Beauty 93}, {\it op. cit.}; M. Chaichian and A. Fridman, {\it Phys. Lett.}
{\bf B298} (1993) 218.
 \item{45.} M. Gronau, A. Nippe and J. L. Rosner, {\it Phys. Rev.} {\bf D47}
(1993) 1988; M. Gronau and J.L. Rosner, in {\it Proceedings of the Workshop on
$B$ Physics at Hadron Accelerators}, Snowmass, CO, June 21 - July 2, 1993, ed.
P. McBride and C. S. Mishra, Fermilab Report FERMILAB-CONF-93/267 (Fermilab,
Batavia, IL, 1993), p.701; {\it Phys. Rev.} {\bf D49} (1994) 254.
\item{46.} C. T. Hill, in {\it Proceedings of the Workshop on
$B$ Physics at Hadron Accelerators}, {\it op. cit}, p.127; C. Quigg, {\it
ibid.}
p.443; E. Eichten, C. T. Hill and C. Quigg, {\it Phys. Rev. Lett.} {\bf 71}
(1994) 4116.
\item{47.} S. Nussinov, {\it Phys. Rev. Lett.} {\bf 35} (1975) 1672.
\item{48.} M. Gronau and J. L. Rosner, {\it Phys. Rev. Lett.} {\bf 72}
(1994) 195.
\item{49.} A. G. Cohen, D. B. Kaplan and A. E. Nelson, {\it Ann. Rev. Nucl.
Part. Sci.} {\bf 43} (1993) 27.

\vfill \eject
\bye